# NONCOMMUTATIVE CHERN-SIMONS SOLITON

Subir Ghosh

Physics and Applied Mathematics Unit,
Indian Statistical Institute,
203 B. T. Road, Calcutta 700108,
India.

Abstract:

We have studied the noncommutative extension of the relativistic Chern-Simons-Higgs model, in the first non-trivial order in $\theta$, with only spatial noncommutativity. Both Lagrangian and Hamiltonian formulations of the problem have been discussed, with the focus being on the canonical and symmetric forms of the energy-momentum tensor. In the Hamiltonian scheme, constraint analysis and the induced Dirac brackets have been provided. The spacetime translation generators and their actions on the fields are discussed in detail.

The effects of noncommutativity on the soliton solutions have been analysed thoroughly and we have come up with some interesting observations. Considering the *relative* strength of the noncommutative effects, we have shown that there is a universal character in the noncommutative correction to the magnetic field - it depends *only* on $\theta$. On the other hand, in the cases of all other observables of physical interest, such as the potential profile, soliton mass or the electric field, $\theta$ as well as $\tau$, (comprising solely of commutative Chern-Simons-Higgs model parameters), appear on similar footings. This phenomenon is a new finding which has come up in the present analysis.

Lastly, we have pointed out a generic problem in the NC extension of the models, in the form of a mismatch between the BPS dynamical equation and the full variational equations of motion, to $O(\theta)$. This mismatch indicates that the analysis is not complete as it brings in to fore the ambiguities in the definition of the energy-momentum tensor in a noncommutative theory.

## Introduction

The Chern-Simons electrodynamics has created a lot of interest in the past. Here the gauge-field dynamics is governed solely by the Chern-Simons term. The gauge theoretic part of this truncation can be perceived as the $\mu \to \infty$ limit of the topologically massive model [1],

$$L_{top} = -\frac{1}{4}F^{\mu\nu}F_{\mu\nu} + \frac{\mu}{4}\epsilon^{\alpha\beta\gamma}F_{\alpha\beta}A_{\gamma}, \qquad (1)$$

$F_{\alpha\beta} = \partial_\alpha A_\beta - \partial_\beta A_\alpha$ being the abelian field tensor. Physically this is realizable in the context of large distance or low energy scales where the Chern-Simons term, with lower number of derivatives dominates over the higher derivative Maxwell term. The charged scalar field, minimally coupled to $U(1)$ Chern-Simons gauge field, with a Higgs type of polynomial potential, gives rise to the celebrated Chern-Simons vortices [2, 3]. For a particular potential profile, one gets a self-dual set of BPS equations where at the self-dual point, the solitons saturate the energy lower bound. These solitons have played significant roles in the context of anyonic quantum field theories [4].

After the relevance of Non-Commutative (NC) quantum field theories [5] was established in the context of string theory [6], soliton solutions in NC theory have generated a great amount of interest. Of the various types of NC solitons discussed so far [7, 8, 9, 10, 11], in the small $\theta$ regime, ($\theta$ being the NC parameter), some NC solitons reach smoothly their corresponding commutative soliton limit [8, 9, 10, 11], whereas in some cases [7], the NC soliton solution ceases to exist in the $\theta \to 0$ limit. The importance of the Chern-Simons theory in commutative spacetime [4] has led to investigations with the NC generalization of the Chern-Simons term [12]. Solitons in the NC Chern-Simons-Higgs system have also been studied in the operatorial framework [9].

In the present work, we will analyze the BPS self-dual solitons of the NC (relativistic) Chern-Simons-Higgs model in a field theoretic framework. For $\theta \to 0$ one recovers the commutative spacetime solitons [2]. The method adapted here was exploited successfully by us in [11], (see also [13]), in the context of NC $CP(1)$ solitons [10]. The scheme utilises the Seiberg-Witten map [6] to convert the NC action to an equivalent action in commutative spacetime, comprising of ordinary field variables. The Seiberg-Witten map is crucial here since the theories in question are gauge theories in ordinary and NC spacetime. All the results derived here are valid to the first non-trivial order in $\theta$. (This is mainly because there are some non-uniqueness in the Seiberg-Witten map in higher orders in $\theta$.) Bogomolnyi analysis of the energy functional reveals the NC BPS solitons which reduce smoothly in the $\theta \to 0$ to their commutative counterpart [2].

We study in detail various features of the NC solitons and come up with some surprising observations. Principal among them is the remarkable fact that, (when the observables are suitably scaled), the $O(\theta)$ correction in the magnetic field (of the soliton) depends *only on $\theta$ and on no other parameters of the theory.* This indicates a sort of universality in the first order NC correction of the magnetic field, at least in these types of planar models. This is in contrast to the other characteristic features of the theory, such as the self-dual potential profile, the electric field or soliton energy, where along with $\theta$ another parameter $\tau$, (comprising solely of the commutative Chern-Simons-Higgs model parameters), plays an equally important role. Since as such $\tau$ is not restricted it is possible to have $\tau$ quite large so that the product $\theta\tau$ is not that small. On the other hand, as we will show, the freedom of choosing $\tau$ can be curbed somewhat via the requirement of the correct nature of the (modified Higgs) potential that can sustain soliton solutions.

Lastly we point out a disagreement between the BPS equations and the full variational equations of motion. This problem cropped up previously in our analysis [11] of the $CP(1)$ soliton. Even though the overall $O(\theta)$ corrected soliton behaves in a smooth and consistent way, we find that to $O(\theta)$ there is a mismatch between the BPS equations and the variational equation of motion. This problem appears to be generic and serious since it puts a question mark on the correctness of the conventional definitions of the Energy Momentum Tensor (EMT) [14], from which the BPS equations originate. As the Bogomolnyi analysis is quite unambiguous in the present case, the above problem can only mean that the energy functional that is minimized in the process is *not* completely correct. We have studied both the canonical and symmetric forms of the EMT and have shown that in the Hamiltonian framework, the correct spacetime translation generators are reproduced from the canonical EMT. These subsidiary checks ensure that the $O(\theta)$ extended model is otherwise a consistent field theory.

The paper is organized as follows: The NC $O(\theta)$ modified version of the Chern-Simons-Higgs model is introduced in Section **II**. The dynamical equations are derived in Section **III**. Section **IV** discusses the canonical EMT in a Hamiltonian framework. Section **V** deals with the constraint analysis and translation generators. Section **VI** comprises computation of the symmetric EMT. Section **VII** is devoted to the Bogomolnyi analysis and BPS equations. We have demonstrated the various NC effects pictorially in Section **VIII**. In Section **IX** we note the above mentioned mismatch between the BPS and variational equations of motion. The paper ends with a conclusion in Section **X**.

## II. Noncommutative Chern-Simons-Higgs model

The spacetime noncommutativity is governed by,

$$[x^\rho, x^\sigma]_* = i\theta^{\rho\sigma}. \tag{2}$$

We restrict ourselves to only spatial noncommutativity ($\theta^{0i} = 0$) and the results are valid to the first non-trivial order in $\theta^{\mu\nu}$. The NC effect is encoded in the replacement of product of functions (in the action) by the associative $*$-product, given by the Moyal-Weyl formula,

$$p(x) * q(x) = pq + \frac{i}{2}\theta^{\rho\sigma}\partial_\rho p \partial_\sigma q + O(\theta^2). \tag{3}$$

The reason for invoking $\theta^{0i} = 0$ is that space-time noncommutativity can induce higher order time derivatives leading to a loss of causality in the field theory [15]. Also, even to $O(\theta)$, it can alter the symplectic structure in a significant way, that might result in a non-perturbative change in the dynamics, which we want to avoid.

In ordinary (commutative) spacetime, the Chern-Simons-Higgs model [1] is described by the Lagrangian,

$$\mathcal{L} = \frac{\mu}{2}\epsilon^{\mu\nu\lambda}A_\mu\partial_\nu A_\lambda + \frac{1}{2} \mid D\phi \mid^2 - V(\mid \phi \mid^2) \tag{4}$$

where $D_\mu\phi = \partial_\phi - ieA_\mu\phi$. We will follow the procedure discussed in [3] where the form of $V(\mid \phi \mid^2)$ is kept arbitrary. It turns out that the BPS equations force it to be of a particular form. We will not repeat the derivation of the well-known soliton solutions of the commutative spacetime [2, 3] as they can be read off from our results by simply putting $\theta = 0$ (the limit being smooth). The NC counterpart of the above model is,

$$\hat{\mathcal{L}} = \frac{\mu}{2}\epsilon^{\mu\nu\lambda}(\hat{A}_\mu * \partial_\nu\hat{A}_\lambda + \frac{2}{3}i\hat{A}_\mu * \hat{A}_\nu * \hat{A}_\lambda) + \frac{1}{2}(\hat{D}^\mu\hat{\phi})^* * \hat{D}_\mu\hat{\phi} - V(\mid \phi \mid^2), \tag{5}$$

[1]Our metric is $g^{\mu\nu} = diag(1, -1, -1)$.

where $\hat{D}_\mu\hat{\phi} \equiv \partial_\mu\hat{\phi} - ie\hat{A}_\mu * \hat{\phi}$. Notice that the products in (4) are replaced by the $*$-product and the NC generalization of the Chern-Simons term is derived in [12]. $\hat{\psi}(x)$ is the NC counterpart of any generic field $\psi(x)$.

Exploiting the Seiberg-Witten map [6, 16] to the first non-trivial order in $\theta$,

$$\hat{\phi} = \phi - \frac{e}{2}\theta^{\alpha\beta}A_\alpha\partial_\beta\phi$$

$$\hat{A}_\mu = A_\mu + \theta^{\sigma\rho}A_\rho(\partial_\sigma A_\mu - \frac{1}{2}\partial_\mu A_\sigma), \quad (6)$$

we recover the $O(\theta)$ corrected Lagrangian for the NC Chern-Simons-Higgs model,

$$\hat{\mathcal{L}} = \frac{\mu}{2}\epsilon^{\mu\nu\lambda}A_\mu\partial_\nu A_\lambda + \frac{1}{2}[\mid D\phi\mid^2 + \frac{e}{2}\theta^{\alpha\beta}\{F_{\alpha\mu}(D_\beta\phi^* D^\mu\phi + D^\mu\phi^* D_\beta\phi)$$

$$-\frac{1}{2}F_{\alpha\beta}\mid D\phi\mid^2\}] - V(\mid\phi\mid^2). \quad (7)$$

A simplified notation is used where $(D_\mu\phi)^* \equiv D_\mu\phi^* = (\partial_\mu + ieA_\mu)\phi^*$ and $F_{\mu\nu} = \partial_\mu A_\nu - \partial_\nu A_\mu$ is the abelian electromagnetic tensor.

We point out a feature of the particular NC extension (5) of the original model of (4). Notice that except the potential term, the rest of the terms in (4) are generalized to NC spacetime in the usual way and once again the form of $V$ is kept arbitrary. As it turns out, the Bogomolny analysis will reveal that the NC extension of the potential is *not* obtainable from the original (commutative spacetime) Higgs potential of [2] via the Seiberg-Witten map.

## III. Equations of motion

The Euler-Lagrange equations of motion for a generic field $\psi_\alpha$,

$$\partial_\mu\frac{\delta\hat{\mathcal{L}}}{\delta(\partial_\mu\psi_\alpha)} - \frac{\delta\hat{\mathcal{L}}}{\delta\psi_\alpha} = 0. \quad (8)$$

yields the following dynamical equations for the model in question (7),

$$\frac{\mu}{2}\epsilon^{\sigma\alpha\beta}F_{\alpha\beta} + \frac{ie}{2}(\phi^* D^\sigma\phi - \phi D^\sigma\phi^*) = \frac{e}{4}\partial_\rho[\theta^{\rho\beta}(D_\beta\phi^* D^\sigma\phi + D^\sigma\phi^* D_\beta\phi) - \theta^{\sigma\beta}(D_\beta\phi^* D^\rho\phi + D^\rho\phi^* D_\beta\phi)$$

$$-\theta^{\rho\sigma}\mid D\phi\mid^2] - \frac{ie^2}{4}[\theta^{\alpha\sigma}F_{\alpha\mu}(\phi^* D^\mu\phi - \phi D^\mu\phi^*) + \theta^{\alpha\beta}F_\alpha{}^\sigma(\phi^* D_\beta\phi - \phi D_\beta\phi^*) - \frac{1}{2}\theta^{\alpha\beta}F_{\alpha\beta}(\phi^* D^\sigma\phi - \phi D^\sigma\phi^*)], \quad (9)$$

$$\frac{1}{2}D_\mu D^\mu\phi + \frac{\delta V}{\delta\phi^*} = -\frac{e}{4}\partial_\rho[\theta^{\alpha\rho}F_{\alpha\mu}D^\mu\phi + \theta^{\alpha\beta}F_\alpha{}^\rho D_\beta\phi] + i\frac{e^2}{4}\theta^{\alpha\beta}[F_{\alpha\mu}(A_\beta D^\mu\phi + A^\mu D_\beta\phi)$$

$$-\frac{1}{2}F_{\alpha\beta}A_\mu D^\mu\phi]. \quad (10)$$

We now restrict ourselves to only spatial noncommutativity, *i.e.* $\theta_{0i} = 0$ and define $\theta_{ij} \equiv \epsilon_{ij}\theta$ , $F_{ij} \equiv \epsilon_{ij}F$. This leads us to the following (manifestly) non-covariant equations corresponding to (9),

$$\mu F = -\frac{ie}{2}(\phi^* D_0\phi - \phi D_0\phi^*)(1 - \frac{e\theta F}{2}) + \frac{e\theta}{4}\epsilon_{ij}\{\partial_i(D_j\phi^* D_0\phi + D_j\phi D_0\phi^*)$$

$$-ieF_{i0}(\phi^* D_j\phi - \phi D_j\phi^*)\} \quad (11)$$

$$\mu\epsilon_{ij}F_{j0} = \frac{ie}{2}(\phi^* D_i\phi - \phi D_i\phi^*)(1+\frac{e\theta F}{2}) + \frac{e\theta}{4}\epsilon_{ij}\{-\partial_0(D_j\phi^* D_0\phi + D_j\phi D_0\phi^*)$$
$$+\partial_j(D_0\phi^* D_0\phi) + ieF_{j0}(\phi^* D_0\phi - \phi D_0\phi^*)\}. \quad (12)$$

We will often use the above equations in the form,

$$\mu F \approx -\frac{ie}{2}(\phi^* D_0\phi - \phi D_0\phi^*) + O(\theta), \quad (13)$$

$$\mu\epsilon_{ij}F_{j0} \approx \frac{ie}{2}(\phi^* D_i\phi - \phi D_i\phi^*) + O(\theta). \quad (14)$$

## IV. Hamiltonian analysis and canonical energy momentum tensor

Let us now introduce the Hamiltonian formulation of the problem at hand. Our aim is to obtain the spacetime and gauge symmetry generators and subsequently study the (spacetime and gauge) transformation properties of the fields. Similar kind of analysis has been done for the $CP(1)$ model in [13]. This requires the construction of the canonical EMT,

$$T_c^{\rho\nu} \equiv \frac{\delta\hat{\mathcal{L}}}{\delta(\partial_\rho A_\sigma)}\partial^\nu A_\sigma + \frac{\delta\hat{\mathcal{L}}}{\delta(\partial_\rho\phi)}\partial^\nu\phi + \frac{\delta\hat{\mathcal{L}}}{\delta(\partial_\rho\phi^*)}\partial^\nu\phi^*. \quad (15)$$

In the present case, we get the canonical EMT

$$T_c^{\rho\nu} = [\frac{\mu}{2}\epsilon^{\mu\rho\sigma}A_\mu + \frac{e}{4}\{\theta^{\rho\beta}(D_\beta\phi^* D^\sigma\phi + D^\sigma\phi^* D_\beta\phi) - \theta^{\sigma\beta}(D_\beta\phi^* D^\rho\phi + D^\rho\phi^* D_\beta\phi)$$
$$-\theta^{\rho\sigma} \mid D\phi \mid^2\}]\partial^\nu A_\sigma + \xi^{\rho\nu} + (\xi^{\rho\nu})^* - g^{\rho\nu}\hat{\mathcal{L}}, \quad (16)$$

where

$$\xi^{\rho\nu} = [\frac{1}{2}(1-\frac{e\theta F}{2})D^\rho\phi^* + \frac{e}{4}(\theta^{\alpha\rho}F_\alpha{}^\beta D_\beta\phi^* + \theta^{\alpha\beta}F_\alpha{}^\rho D_\beta\phi^*)]\partial^\nu\phi.$$

The energy and momentum densities follow immedietly:

$$T_{00}^c = V - \mu A_0 F + \frac{1}{2}(1-\frac{e\theta F}{2}) \mid D_0\phi \mid^2 + \frac{1}{2}(1+\frac{e\theta F}{2}) \mid D_k\phi \mid^2 + \frac{ie}{2}(1-\frac{e\theta F}{2})A_0(\phi D_0\phi^* - \phi^* D_0\phi)$$
$$+\frac{e\theta}{4}\epsilon_{kj}[F_{k0}\{D_j\phi^* D_0\phi + D_j\phi D_0\phi^* + ieA_0(\phi D_j\phi^* - \phi^* D_j\phi)\} - \partial_k A_0(D_j\phi^* D_0\phi + D_j\phi D_0\phi^*)], \quad (17)$$

$$T_c^{0i} = \frac{\mu}{2}\epsilon_{jk}A_j\partial_i A_k - \frac{1}{2}(1-\frac{e\theta F}{2})(D_0\phi^*\partial_i\phi + D_0\phi\partial_i\phi^*)$$
$$-\frac{e\theta}{4}\epsilon_{jk}(D_j\phi^* D_0\phi + D_j\phi D_0\phi^*)\partial_i A_k$$
$$-\frac{e\theta}{4}\epsilon_{jk}F_{k0}(D_k\phi^*\partial_i\phi + D_k\phi\partial_i\phi^*). \quad (18)$$

The next task is to introduce the canonical momenta which will indicate that the theory has constraints and hence a Hamiltonian constraint analysis becomes imperative. Defining the momenta as,

$$\pi^* \equiv \frac{\delta \mathcal{L}}{\delta \dot{\phi}^*}\ ,\ \pi \equiv \frac{\delta \mathcal{L}}{\delta \dot{\phi}}\ ,\ \pi^\mu \equiv \frac{\delta \mathcal{L}}{\delta \dot{A}_\mu}$$

we find,

$$\pi^* = \frac{1}{2}(1 - \frac{e\theta}{2})D_0\phi + \frac{e\theta}{4}\epsilon_{ij}F_{i0}D_j\phi \approx \frac{1}{2}D_0\phi + O(\theta),$$

$$\pi = \frac{1}{2}(1 - \frac{e\theta}{2})D_0\phi^* + \frac{e\theta}{4}\epsilon_{ij}F_{i0}D_j\phi^* \approx \frac{1}{2}D_0\phi^* + O(\theta),$$

$$\pi_0 = 0\ ;\ \pi_k = \frac{\mu}{2}\epsilon_{kj}A_j - \frac{e\theta}{4}\epsilon_{kj}(D_j\phi^* D_0\phi + D_j\phi D_0\phi^*) \approx \frac{\mu}{2}\epsilon_{kj}A_j + O(\theta). \tag{19}$$

This allows us to rewrite the above defining relations in (19) to $O(\theta)$ in the following way:

$$\pi^* = \frac{1}{2}D_0\phi - \frac{e\theta}{2}F\pi^* + \frac{e\theta}{4}\epsilon_{ij}F_{i0}D_j\phi + O(\theta^2),$$

$$\pi = \frac{1}{2}D_0\phi^* - \frac{e\theta}{2}F\pi + \frac{e\theta}{4}\epsilon_{ij}F_{i0}D_j\phi^* + O(\theta^2), \tag{20}$$

$$\pi_k = \frac{\mu}{2}\epsilon_{kj}A_j - \frac{e\theta}{2}\epsilon_{kj}(\pi^* D_j\phi^* + \pi D_j\phi) + O(\theta^2). \tag{21}$$

We also invert the above relations to $O(\theta)$ and get,

$$D_0\phi = 2\pi^* + e\theta(F\pi^* - \frac{1}{2}\epsilon_{ij}F_{i0}D_j\phi) + O(\theta^2),$$

$$D_0\phi^* = 2\pi + e\theta(F\pi - \frac{1}{2}\epsilon_{ij}F_{i0}D_j\phi^*) + O(\theta^2). \tag{22}$$

Reexpressed in terms of the phase space variable the total energy and momenta look like,

$$H_c = \int d^2x\ T_c^{00} = \int d^2x\ [2(1 + \frac{e\theta F}{2})(\pi^*\pi + \frac{1}{4}D_i\phi^* D_i\phi) + V + A_0\mathcal{G}], \tag{23}$$

$$P_i^c = \int d^2x\ T_{0i}^c = \int d^2x\ [\pi\partial_i\phi + \pi^*\partial_i\phi^* + \pi_j\partial_i A_j)], \tag{24}$$

where

$$\mathcal{G} \equiv -\partial_i\pi_i - \frac{\mu}{2}F + ie(\phi\pi - \phi^*\pi^*) \approx 0 \tag{25}$$

is the Gauss law constraint besides the trivial one $\pi_0 \approx 0$.

## V. Constraints, Dirac brackets and Translation generators

The two constraints, $\mathcal{G} \approx 0$ and $\pi_0 \approx 0$, constitute the First Class Constraints (FCCs) of the present theory in the terminology of Dirac [17]. The FCCs commute (in the sense of Poisson brackets) and signify local gauge invariance (which is $U(1)$ in the present case).

Besides the above FCCs, following from the relation (21) there are furthermore two Second Class Constraints (SCC) [17],

$$\chi_i \equiv \pi_i - \frac{\mu}{2}\epsilon_{ij}A_j(1-P) + \frac{e\theta}{2}\epsilon_{ij}(\pi\partial_j\phi + \pi^*\partial_j\phi^*), \tag{26}$$

where

$$P \equiv -\frac{i\theta e^2}{\mu}(\phi\pi - \phi^*\pi^*).$$

The SCCs are non-commuting (in the sense of Poisson brackets) and they induce a change in the symplectic structure, whereby a generic Poisson bracket ($\{A, B\}$) is replaced by Dirac a bracket ($\{A, B\}_{DB}$), defined in the following way,

$$\{A, B\}_{DB} = \{A, B\} - \{A, \chi_i\}\chi_{ij}^{-1}\{\chi_j, B\}. \tag{27}$$

In the above definition (27), $\chi_{ij}^{-1}$ denotes the inverse of the Poisson bracket matrix,

$$\chi_{ij}(x, Y) \equiv \{\chi_i(x), \chi_j(y)\} = -\mu\epsilon_{ij}(1-P)\delta(x-y), \tag{28}$$

where $P = \frac{i\theta e^2}{\mu}(\pi^*\phi^* - \pi\phi)$. The inverse is computed to be,

$$\chi_{jk}(x, y)^{-1} = \frac{1}{\mu}\epsilon_{jk}(1+P)\delta(x-y) + O(\theta^2). \tag{29}$$

Utilising the defining equation (27), it is now straightforward to obtain the full set of Dirac brackets which are given below:

$$\{A_i(x), A_j(y)\} = \frac{1}{\mu}\epsilon_{ij}(1+P)\delta(x-y)\ ;\ \{A_i(x), \pi_j(y)\} = \frac{1}{2}\delta_{ij}\delta(x-y)\ ;$$

$$\{A_i(x), \phi(y)\} = -\frac{e\theta}{2\mu}D_i\phi\delta(x-y)\ ;\ \{A_i(x), \pi(y)\} = \frac{e\theta}{2\mu}\pi(x)D_i^{(x)}\delta(x-y)$$

$$\{\pi_i(x), \pi_j(y)\} = \frac{\mu}{4}(1-P)\epsilon_{ij}\delta(x-y)\ ;\ \{\pi_i(x), \phi(y)\} = \frac{e\theta}{4}\epsilon_{ij}D_j\phi\delta(x-y)\ ;$$

$$\{\pi_i(x), \pi(y)\} = -\frac{e\theta}{4}\epsilon_{ij}\pi(x)D_j^{(x)}\delta(x-y), \tag{30}$$

$$\{\phi(x), \pi(y)\} = \delta(x-y) + O(\theta^2)\ ;\ \{\phi(x), \phi(y)\} = \{\pi(x), \pi(y)\} = O(\theta^2). \tag{31}$$

It should be remembered that *all* the above relations are valid up to $O(\theta)$. Notice that in this approximation, there is no modification in the $\phi - \pi$ sector. Also note that, starting from the set of relations (30) above, and in the subsequent discussion, unless otherwise stated, *all* the brackets are Dirac brackets and so we have dropped the subscript $\{,\}_{DB}$ from now on.

Our first objective is to apply the Dirac brackets to ensure that the fields are transforming in the proper way under the symmetry transformations. We start with gauge transformation, the infinitesimal transformation generator of which is given by,

$$G \equiv \int d^2x\ \lambda(x)\mathcal{G}(x), \tag{32}$$

$\lambda(x)$ being the infinitesimal parameter. Utilising the Dirac brackets (30,31) we find,

$$\{\phi(x), \mathcal{G}(y)\} = ie\phi(x)\delta(x-y) \ \rightarrow \ \{\phi(x), G\} = ie\lambda(x)\phi(x), \quad (33)$$

$$\{A_i(x), \mathcal{G}(y)\} = \partial_i^{(x)}\delta(x-y) \ \rightarrow \ \{A_i(x), G\} = \partial_i\lambda(x). \quad (34)$$

Thus the gauge properties of the charged scalar field $\phi$ and the $U(1)$ gauge field $A_i$ are correctly recovered.

Next we study the spacetime transformation properties of the fields. It is now a simple task to establish the following relation,

$$\{\psi(x), P_i^c\} = \partial_i\psi(x), \quad (35)$$

where $\psi \equiv \{\phi, \pi, A_i, \pi_i\}$. The expression for $P_i^c$ is given in (24) and the new symplectic structure (30, 31) is used. This indicates that the momentum operator $P_i^c$ correctly plays the role of the generator of spatial translation. From the explicit form of $P_i^c$ given in (24) it is evident that the translation generator is essentially canonical and that the noncommutativity has not generated any extra contribution. This feature obviously reflects the translation invariance of the starting model (7). An identical situation prevailed in [13].

However, recovering the Hamiltonian form of the equations of motion, (which is equivalent to obtaining the time derivatives correctly), turns out to be somewhat non-trivial. The following bracket,

$$\{\phi(x), H^c\} = D_0\phi(x) \quad (36)$$

indicates that the proper definition of time derivative for $\phi$ *i.e.*

$$\{\phi(x), H^c\} = \partial_0\phi(x) \quad (37)$$

requires a gauge fixing $A_0 \approx 0$. Indeed, this gauge choice is harmless as far as the Dirac brackets are concerned since it simply removes the SCCs $\pi_0$ and $A_0$, (that constitutes a canonically conjugate pair), from further considerations without affecting rest of the brackets. Considering $A_i$ we find,

$$\{A_i(x), H^c\} = -F_{i0}(1+\frac{e\theta}{\mu}\mathcal{G}) - \frac{e\theta}{4\mu}[D_i\phi^* D_\mu D^\mu\phi + D_i\phi D_\mu^*(D^\mu\phi)^* - \partial_i(\partial_0\phi^*\partial_0\phi)] - \frac{e\theta}{2\mu}\partial_i V. \quad (38)$$

In deriving the above relation, we have used $A_0 = 0$ gauge. Exploiting the dynamical equation for $\phi$ from (10),

$$D_\mu D^\mu\phi = -2\frac{\delta V}{\delta\phi^*} + O(\theta) \ ; \ D_\mu^*(D^\mu\phi)^* = -2\frac{\delta V}{\delta\phi} + O(\theta),$$

in the above equation, we find a simplified relation,

$$\{A_i(x), H^c\} = \partial_0 A_i + \frac{e\theta}{2\mu}\partial_i(\partial_0\phi^*\partial_0\phi) + O(\theta^2). \quad (39)$$

Notice that there still remains an $O(\theta)$ extra piece. However it has the structure of a $U()1$ gauge transformation. Since the Gauss law FCC is still intact, we are allowed to make a further gauge transformation thus maintaining the proper definition of a time derivative. This feature is reminiscent of gauge theories in commutative spacetime, where the gauge field $A_i$ behaves properly under Lorentz boosts modulo a gauge transformation. It should be remembered that in our study [13] of the $CP(1)$ model also, deriving the Hamiltonian equations of motion turned out to be more involved.

## VI. Symmetric energy momentum tensor

Let us now construct the symmetric form of the EMT that is to be utilised in obtaining the BPS soliton solutions of the NC Chern-Simons-Higgs model. This has been the common practice in existing literature for commutative [2, 3] and noncommutative [9] Chern-Simons-Higgs solitons. The symmetric form of the EMT is conventionally defined as [14],

$$T^s_{\mu\nu} \equiv \frac{2}{\sqrt{-g}}\frac{\delta \mathcal{L}}{\delta g^{\mu\nu}}. \tag{40}$$

In the present case this is obtained by coupling the model with the metric field $g_{\mu\nu}$ (with $g_{\mu\nu}$ being a background field) and finally reducing it to a flat Minkowski background. This generates the symmetric EMT,

$$\begin{aligned} T^s_{\mu\nu} &= -g_{\mu\nu}\bar{\mathcal{L}} + \frac{1}{2}(D_\mu\phi^* D_\nu\phi + D_\nu\phi^* D_\mu\phi)(1 - \frac{e\theta F}{2}) - \frac{e}{4}D_\beta\phi^* D^\beta\phi(\theta_{\mu\alpha}F_\nu{}^\alpha + \theta_{\nu\alpha}F_\mu{}^\alpha) \\ &+\frac{e}{4}[(\theta_{\mu\alpha}F_{\nu\beta} + \theta_{\nu\alpha}F_{\mu\beta})(D^\alpha\phi^* D^\beta\phi + D^\beta\phi^* D^\alpha\phi) + F^{\alpha\beta}\{\theta_{\alpha\mu}(D_\nu\phi^* D_\beta\phi + D_\beta\phi^* D_\nu\phi) \\ &+\theta_{\alpha\nu}(D_\mu\phi^* D_\beta\phi + D_\beta\phi^* D_\mu\phi)\} + \theta^{\alpha\beta}\{F_{\alpha\mu}(D_\beta\phi^* D_\nu\phi + D_\nu\phi^* D_\beta\phi) + F_{\alpha\nu}(D_\beta\phi^* D_\mu\phi + D_\mu\phi^* D_\beta\phi)\}]. \end{aligned} \tag{41}$$

In the above expression, $\bar{\mathcal{L}}$ represents the Lagrangian (7) without the Chern-Simons term since the topological term is metric independent and does not contribute in the variation of the metric tensor.

It is straightforward to check that for $\theta = 0$, the symmetric EMT is conserved,

$$\partial^\mu T^s_{\mu\nu} \mid_{\theta=0} = 0.$$

Although the same is not true for the $O(\theta)$ corrected $T^s_{\mu\nu}$, we have explicitly checked that a modified EMT can be defined in the present case which is symmetric and conserved (see Das and Frenkel in [14]).

Now (41) leads to the following expression for the energy density,

$$\begin{aligned} T^s_{00} &= \frac{1}{2}(1-\alpha)D_0\phi^* D_0\phi + \frac{1}{2}(1+\alpha)D_i\phi^* D_i\phi + \frac{e\theta}{4}\epsilon_{ij}F_{i0}(D_j\phi^* D_0\phi + D_j\phi D_0\phi^*) + V \\ &= \frac{1}{2}(1-\alpha)D_0\phi^* D_0\phi + \frac{1}{2}(1+\alpha)D_i\phi^* D_i\phi \\ &-\frac{i\theta e^2}{8\mu}(\phi^* D_j\phi - \phi D_j\phi^*)(D_j\phi^* D_0\phi + D_j\phi D_0\phi^*) + V, \end{aligned} \tag{42}$$

where $\alpha = \frac{e\theta F}{2}$ and the equations of motion (12) have been used.

As a curiosity, let us express the Hamiltonian and momenta obtained from the symmetric EMT in terms of the phase space variables defined before in (19). We find

$$\int d^2x \; T^{00}_s = \int d^2x \; [2(1 + \frac{e\theta F}{2})(\pi^*\pi + \frac{1}{4}D_i\phi^* D_i\phi) + V], \tag{43}$$

which agrees with the canonical form $H_c$ given in (23) on the constraint surface. The expression for the (symmetric) momentum density as a function of phase space degrees of freedom is,

$$T^s_{0i} = T^s_{i0} = (\pi D_i\phi + \pi^* D_i\phi^*)(1 + e\theta F) + \frac{i\theta e^2}{8\mu}[4(\phi^* D_i\phi - \phi D_i\phi^*)(\mid \pi \mid^2$$

$$+\frac{1}{4} \mid D_i\phi \mid^2) + ((\phi^* D_j\phi - \phi D_j\phi^*)(D_j\phi^* D_i\phi + D_j\phi D_i\phi^*). \tag{44}$$

This is quite distinct from the canonical expression of momentum obtained in (24).

In this context let us note a puzzling feature regarding the conservation of the energy. We have shown in (43) that in terms of phase space variables, the energy obtained from the symmetric EMT agrees with the energy obtained from the canonical EMT. In the previous section, we have also shown that this Hamiltonian correctly generates time translations and hence it should be conserved as well. This does not appear to be valid in the Lagrangian picture.

## VII. Bogomolnyi analysis and BPS equations for soliton

Our approach is same as that of the commutative spacetime Bogomolnyi analysis [2, 3]. Similar analysis for the NC $CP(1)$ model was performed in [11]. We concentrate on the energy expression provided in (42) and write it in the form,

$$H^s = \int d^2x\, T^s_{00}(x) = \int \frac{1}{2}(1+\alpha) \mid D_\pm\phi \mid^2 + \frac{1}{2} \mid \sqrt{1-\alpha}D_0\phi \pm \frac{ie^2}{2\mu}\sqrt{1+\alpha}(\mid \phi \mid^2 - v^2)\phi - \frac{\theta e^2}{4\mu}\gamma_j D_j\phi \mid^2$$

$$\mp\frac{ev^2}{2\mu}J_0 \pm \frac{e\alpha}{2\mu}(\mid \phi \mid^2 - v^2)J_0 + V - \frac{e^4}{8\mu^2}(1+\alpha)(\mid \phi \mid^2 - v^2)^2 \mid \phi \mid^2$$

$$\mp\frac{1}{2}(e\alpha\epsilon_{ij}A_j\partial_i \mid \phi \mid^2 + i\alpha\epsilon_{ij}\partial_i\phi^*\partial_j\phi) \mp \frac{\theta e^4}{16\mu^2}(\mid \phi \mid^2 - v^2)(\phi^* D_j\phi - \phi D_j\phi^*)^2$$

$$\mp\frac{i\theta e^4}{16\mu^2}(\mid \phi \mid^2 - v^2)(\phi^* D_j\phi - \phi D_j\phi^*)\gamma_j, \tag{45}$$

where,

$$D_\pm \equiv D_1 \pm iD_2\ ;\ \gamma_j \equiv \pm\epsilon_{ij}\partial_i \mid \phi \mid^2 + i(\phi^* D_j\phi - \phi D_j\phi^*).$$

The following identities have been exploited:

$$\int d^2x\ \frac{1}{2}D_i\phi^* D_i\phi = \int d^2x\ [\frac{1}{2} \mid D_\pm\phi \mid^2 \pm \frac{e}{2}F \mid \phi \mid^2]$$

$$\int d^2x\ \frac{1}{2}\alpha D_i\phi^* D_i\phi = \int d^2x\ [\alpha(\frac{1}{2} \mid D_\pm\phi \mid^2 \pm \frac{e}{2}F \mid \phi \mid^2) \mp \frac{1}{2}(e\alpha F \mid \phi \mid^2$$

$$+e\alpha\epsilon_{ij}A_j\partial_i \mid \phi \mid^2 + i\alpha\epsilon_{ij}\partial_i\phi^*\partial_j\phi)] \tag{46}$$

Also note that similar to the commutative case [3] we have defined the *conserved* $U(1)$ current from (9) in the form,

$$\epsilon^{\mu\nu\lambda}F_{\nu\lambda} \equiv -\frac{2}{\mu}J^\mu \quad \Rightarrow \partial_\mu J^\mu = 0. \tag{47}$$

This means that our expression for the conserved current contains $O(\theta)$ terms and $J_0$ is the conserved charge density. However, defining the current in this way ensures that the charge-flux equality will still remain intact. In particular, in arriving at (45), we have used the relation,

$$\frac{ie}{2}(\phi^* D_0\phi - \phi D_0\phi^*) = (1+\alpha)J_0 - \frac{\theta e^3}{8\mu}(\phi^* D_j\phi - \phi D_j\phi^*)^2 + \frac{\theta e}{4}\epsilon_{ij}\partial_i(D_j\phi^* D_0\phi + D_j\phi D_0\phi^*). \quad (48)$$

It can be checked that upon simplification, the last but one line of (45) vanishes identically. Let us now make the ansatz that

$$D_\pm\phi = O(\theta) \;\Rightarrow\; D_i\phi = \pm i\epsilon_{ij}D_j\phi + O(\theta). \quad (49)$$

This is justified since we expect $O(\theta)$ modifications in the results pertaining to commutative spacetime ($\theta = 0$). This immediately leads to

$$\gamma_j = O(\theta). \quad (50)$$

Hence one can ignore the terms containing $\gamma_j$ since they are always multiplied by $\theta$. This simplifies the situation considerably and we are led to the cherished BPS equations for the solitons of the NC Chern-Simons-Higgs theory, in the lowest non-trivial order in $\theta$:

$$D_\pm\phi = 0 \Rightarrow D_i\phi = \pm i\epsilon_{ij}D_j\phi\ , \quad (51)$$

$$D_0\phi \pm \frac{ie^2}{2\mu}(1+\alpha)(\mid\phi\mid^2 - v^2)\phi = 0, \quad (52)$$

with $\gamma_j$ vanishing on the BPS shell. This constitutes our main result. The analysis determines the potential profile to be,

$$V = \frac{e^4}{8\mu^2}(1+\alpha)(\mid\phi\mid^2 - v^2)^2 \mid\phi\mid^2\ . \quad (53)$$

Let us choose the *lower* sign in the BPS equations (51,52), which we refer to as the self-dual solution. The upper sign will correspond anti-self-dual solution.

We want to express the gauge field quantities in terms of the $\phi$-fields by exploiting the BPS equations. This is convenient since the BPS equations are first order in derivatives.

First of all, from the equation of motion (11) we obtain $F$ as,

$$F \equiv -\frac{J_0}{\mu} = -\frac{ie}{2\mu}(\phi^* D_0\phi - \phi D_0\phi^*) + \frac{\theta e^3}{8\mu^2}[(\phi^* D_0\phi - \phi D_0\phi^*)^2$$

$$-(\phi^* D_i\phi - \phi D_i\phi^*)^2] + \frac{\theta e}{4\mu}\epsilon_{ij}\partial_i(D_j\phi^* D_0\phi + D_j\phi D_0\phi^*) \quad (54)$$

Substituting the BPS equations in the above relation, we get for the self-dual case,

$$F_{sd} = \frac{e^3}{2\mu^2}(\mid\phi\mid^2 - v^2)\mid\phi\mid^2 - \frac{\theta e^3}{8\mu^2}(\mid\phi\mid^2 - v^2)\nabla\mid\phi\mid^2, \quad (55)$$

where $\nabla \equiv \partial_i\partial_i$. Putting this back in the expression for the energy (45) we get the BPS saturated energy lower bound (or equivalently the soliton mass) as,

$$W_{sd} = \int [\frac{ev^2}{2\mu}J_0 - \frac{\theta e^2}{4\mu^2}(v^2 - \phi^2)J_0^2] = \int [-\frac{ev^2}{2}F - \frac{\theta e^2}{4}(v^2 - \phi^2)F^2]. \tag{56}$$

As a final step, we can express the potential and energy completely in terms of $\phi$:

$$V_{sd} = \frac{e^4}{8\mu^2}[(\mid \phi \mid^2 - v^2)^2 \mid \phi \mid^2 + \theta \frac{e^4}{4\mu^2}(\mid \phi \mid^2 - v^2)^3(\mid \phi \mid^2)^2], \tag{57}$$

$$W_{sd} = \int \frac{e^4v^2}{4\mu^2}[-(\mid \phi \mid^2 - v^2) \mid \phi \mid^2 + \theta\{\frac{e^4}{4\mu^2v^2}(\mid \phi \mid^2 - v^2)^3(\mid \phi \mid^2)^2 + \frac{1}{4}(\mid \phi \mid^2 - v^2)\nabla \mid \phi \mid^2\}]. \tag{58}$$

The electric field is obtained in the form,

$$(E_k)_{sd} \equiv (F_{k0})_{sd} = -\frac{e}{2\mu}\partial_k \mid \phi \mid^2 + \frac{\theta e^5}{16\mu^3}(\mid \phi \mid^2 - v^2)(3 \mid \phi \mid^2 - v^2)\partial_k \mid \phi \mid^2, \tag{59}$$

from which we get the magnitude of the electric field

$$(E_kE_k)_{sd} = \frac{e^2}{4\mu^2}\partial_k \mid \phi \mid^2 \partial_k \mid \phi \mid^2 - \frac{\theta e^6}{16\mu^4}\partial_k \mid \phi \mid^2 \partial_k \mid \phi \mid^2 (\mid \phi \mid^2 - v^2)(3 \mid \phi \mid^2 - v^2). \tag{60}$$

By choosing a simple time dependence of the $\phi$-field [3], such as

$$\phi(x,t) = e^{-i\frac{e^2v^2}{2\mu}t}\varphi(x), \tag{61}$$

$(A_0)_{sd}$ is identified as,

$$(A_0)_{sd} = -\frac{e}{2\mu} \mid \phi \mid^2 [1 + \frac{\theta e^4}{4\mu^2}(\mid \phi \mid^2 - v^2)^2]. \tag{62}$$

Analogous expressions for the anti-self-dual soliton solutions are,

$$V_{asd} = \frac{e^4}{8\mu^2}[(\mid \phi \mid^2 - v^2)^2 \mid \phi \mid^2 - \theta \frac{e^4}{4\mu^2}(\mid \phi \mid^2 - v^2)^3(\mid \phi \mid^2)^2], \tag{63}$$

$$F_{asd} = -\frac{e^3}{2\mu^2}(\mid \phi \mid^2 - v^2) \mid \phi \mid^2 - \frac{\theta e^3}{8\mu^2}(\mid \phi \mid^2 - v^2)\nabla \mid \phi \mid^2, \tag{64}$$

$$(E_kE_k)_{asd} = \frac{e^2}{4\mu^2}\partial_k \mid \phi \mid^2 \partial_k \mid \phi \mid^2 - \frac{\theta e^6}{16\mu^4}\partial_k \mid \phi \mid^2 \partial_k \mid \phi \mid^2 ((\mid \phi \mid^2)^2 - v^4), \tag{65}$$

$$W_{asd} = \int \frac{e^4v^2}{4\mu^2}[-(\mid \phi \mid^2 - v^2) \mid \phi \mid^2 - \theta\{\frac{e^4}{4\mu^2v^2}(\mid \phi \mid^2 - v^2)^3(\mid \phi \mid^2)^2 + \frac{1}{4}(\mid \phi \mid^2 - v^2)\nabla \mid \phi \mid^2\}]. \tag{66}$$

As an aside, it is interesting to observe that, using the Seiberg-Witten map for the $\phi$-variables,

$$\hat{\phi}^*\hat{\phi} = \phi^*\phi + \frac{1}{2}\theta^{\alpha\beta}[i\partial_\alpha\phi^*\partial_\beta\phi + eA_\beta\partial_\alpha \mid \phi \mid^2] + O(\theta^2), \tag{67}$$

it is not possible to correctly generate the $O(\theta)$ term in the potential in (57) from the $\theta$-independent potential term. The latter, being the Higgs potential for commutative spacetime, can be obtained from the NC potential by putting $\theta = 0$. This phenomenon indicates that the NC extension of the Chern-Simons-Higgs soliton is not derivable by just applying the Seiberg Witten map on the commutative model. The potential has to be tuned properly.

## VIII. Effects of noncommutativity on the soliton

We now demonstrate the effects of noncommutativity on the soliton. It is quite remarkable that the electric and magnetic fields of the soliton are affected in very *different* ways. In fact, the parameters of the Chern-Simons-Higgs model ($e$, $\mu$ and $v$) enter in the fray in a non-trivial way. This can be noticed after appropriate scaling of the observables as we now illustrate. There appears to be a universal (*i.e.* parameter independent) nature in the noncommutative effects in the magnetic field as it depends only on $\theta$. However, this property is not shared by the electric field or the potential, where $\theta$ as well as the combination $\tau \equiv \frac{e^4v^4}{8\mu^2}$ plays equally important roles. Note that $\tau$ consists entirely of parameters of the (commutative) Chern-Simons-Higgs model and as such is not restricted by any bounds. However, in presence of noncommutativity, formation of the potential well demands certain bounds on the value of $\tau$. These features were not reported in the earlier literature [9].

We now move on to the axially symmetric solutions where $n$ elementary vortices are superimposed at the origin. As has been discussed before [11, 13], the spatial noncommutativity in 2+1-dimensions does not destroy the rotational symmetry. (This is also evident from the canonical nature of the momentum operator discussed here.) Notice that in (51) and (52), the noncommutativity has modified only part of the full set of BPS equations. We try solutions of the form,

$$\phi = vg(r)e^{in\theta} \ ; \ eA_i = \epsilon_{ij}\frac{\hat{r}_j}{r}[a(r) - n], \tag{68}$$

with $F = -\frac{a'}{er}$. This brings us to the set of consistency condition,

$$-\frac{a'}{er} = \frac{e^3v^4}{2\mu^2}[g^2(g^2-1) - \frac{\theta}{4}(g^2-1)\nabla g^2]. \tag{69}$$

This shows that to linear order in $\theta$ we can use the same expression for $g(r)$ as given in [2] but $a(r)$ requires a $\theta$-correction term, which can also be expressed in terms of $g(r)$. The scenario can be compared with the commutative case [2]. From now on we will exploit only the form of $g(r)$ with the same boundary conditions as given in [2] and consider the single soliton case, *i.e.* $n = 1$.

In *Fig.* I-A, *Fig.* I-B and *Fig.* II we have shown the noncommutativity effect on the potential. We have plotted,

$$V_{sd}(g)/(\frac{e^4v^6}{8\mu^2} \equiv VS(g) = (g^2-1)g^2 + 2\theta\tau(g^2-1)^3g^4,$$

$$V_{asd}(g)/(\frac{e^4v^6}{8\mu^2} \equiv VA(g) = (g^2-1)g^2 - 2\theta\tau(g^2-1)^3g^4, \tag{70}$$

with $V(g)$ representing the $\theta = 0$ case. The points to notice are: (i) The Chern-Simons-Higgs parameter $\tau$ appears in the expressions. (ii) The correction term tends to flatten the potential humps in the self-dual case (see *Fig.* II.)
(iii)For the anti-self-dual case, the well structure can disappear when some critical values of the $\theta\tau$ combination is reached (see *Fig.* I-B). Since $\theta$ is assumed to be small, bounds can be put on the value on the scale of $\tau$.

In *Fig.* III-A, *Fig.* III-B and *Fig.* III-C, we have plotted the magnetic field $F$ as a function of $r$ given by

$$F_{sd}(r)/(\frac{e^3v^4}{2\mu^2}) \equiv BS(r) = [g^2(g^2-1) - \frac{\theta}{4}(g^2-1)\nabla g^2],$$

$$-F_{asd}(r)/(\frac{e^3v^4}{2\mu^2}) \equiv BA(r) = [g^2(g^2-1) + \frac{\theta}{4}(g^2-1)\nabla g^2], \qquad (71)$$

where as before $B(r)$ is the $\theta = 0$ result. Note that $\tau$ does not appear in this relative $F$ profile. The significant facts are: (i) The Chern-Simons-Higgs parameter $\tau$ is absent and only the NC parameter $\theta$ determines the relative strength of the magnetic fields with or without the NC correction.
(ii) The NC effect is not very pronounced for both self-dual and anti-self-dual cases.
The universal nature of the NC correction term in the magnetic field that we mentioned before is clearly visible. [2]

In *Fig.* IV-A, *Fig.* IV-B and *Fig.* IV-C, we plot the radial component of the electric field as a function of $r$ given by,

$$E_{sd}(r)/(\frac{ev^2}{\mu}) \equiv ES(r) = gg'[1 - \theta\tau(g^2-1)(3g^2-1)],$$

$$E_{asd}(r)/(\frac{ev^2}{\mu}) \equiv EA(r) = gg'[1 - \theta\tau(g^4-1)], \qquad (72)$$

with $E(r)$ giving the $\theta = 0$ result. The major points to note are: (i) Both $\theta$ and $\tau$ appear in the expressions.
(ii) The functional form of the correction terms in the self-dual and anti-self-dual cases are quite distinct.

Lastly in *Fig.* V-A, *Fig.* V-B and *Fig.* V-C, we plot the effect of noncommutativity on the energy or equivalently the soliton mass, where

$$-W_{sd}(r)/(\frac{e^4v^6}{4\mu^2}) \equiv WS(r) = g^2(g^2-1) - \theta[\frac{1}{4}(g^2-1)\nabla g^2 + 2\tau g^4(g^2-1)],$$

$$-W_{asd}(r)/(\frac{e^4v^6}{4\mu^2}) \equiv WA(r) = g^2(g^2-1) + \theta[\frac{1}{4}(g^2-1)\nabla g^2 + 2\tau g^4(g^2-1)], \qquad (73)$$

with $W(r)$ representing the $\theta = 0$ result. There are several interesting points to note: (i) Both $\theta$ and $\tau$ enter the picture.

[2] In this context, we would like comment that in the work of Lozano et.al. in [9], the graphs with values of the parameter $a = (v^2\theta)/(2\kappa^2)$, (where $\kappa$ is identified with $\mu$ in our case) chosen as 2, 1 and 0.5, do not faithfully represent the NC effect.

(ii) In the commutative case, the energy was directly proportional to the magnetic field. On the contrary, for non-zero $\theta$, the energy depends as before on the NC magnetic field but there is also the $\tau$-dependent extra contribution,

$$WS(r) \equiv BS - 2\theta\tau g^4(g^2-1),$$
$$WA(r) \equiv BA + 2\theta\tau g^4(g^2-1). \qquad (74)$$

## XI. BPS vs. variational equations of motion

In a previous analysis [11] of the NC $CP(1)$ model, we had commented on the inadequacy of the conventional definitions (*i.e.* canonical or symmetric) of the EMT in the context of NC $CP(1)$ model. We had demonstrated [11] that although the Bogomolnyi analysis of the expression for "energy" generated a set of BPS equations, ( the solution of which revealed the soliton unambiguously), the BPS equations did not completely satisfy the variational equations of motion. Obviously this is needed for consistency of the whole procedure. This mismatch indicates that the standard definition of energy that has been used is not exactly the correct one, at least in the NC models studied in [11] and here. It is important to understand whether this feature is generic to NC field theories. This brings us to a similar consistency check in the present NC Chern-Simons-Higgs theory. We find a similar type of mismatch in the present context as well, indicating that a deeper study of the problem is needed.

Our aim is to check whether the full set of BPS equations (51), (52) and (9) is consistent with (10) - the (second order) variational equation of motion for $\phi$. We assume the validity of (51), (52) and (9) in order to simplify the right hand side of (10) and obtain,

$$\frac{1}{2}D_\mu D^\mu\phi+\frac{\delta V}{\delta\phi^*} = \frac{\theta e}{4}[\frac{e^3}{2\mu^2}D_i\phi\{(\mid\phi\mid^2-v^2)\phi\partial_i\phi^*+2\partial_i((\mid\phi\mid^2-v^2)\mid\phi\mid^2)+2ie(\mid\phi\mid^2-v^2)\mid\phi\mid^2 A_i\}$$
$$+2eF^2\phi - ieFA_0D_0\phi]. \qquad (75)$$

On the other hand, simplifying the left hand side of (10), once again by exploiting the BPS equations ((51), (52)) one can check that there are no $\theta$-independent terms and the remaining $O(\theta)$ term cancels with the last line of the right hand side of (75). This shows that the BPS equations and variational (second order) equations are consistent for the commutative case ($\theta = 0$) case and the mismatch to $O(\theta)$ boils down to the equation,

$$D_i\phi\{(\mid\phi\mid^2-v^2)\phi\partial_i\phi^* + 2\partial_i((\mid\phi\mid^2-v^2)\mid\phi\mid^2) + 2ie(\mid\phi\mid^2-v^2)\mid\phi\mid^2 A_i\} = 0. \qquad (76)$$

One can further utilise the BPS equations (51) and (52), to express $A_i$ in terms of $\phi$ in the following way,

$$A_i = \frac{i}{2e\mid\phi\mid^2}(i\epsilon_{ij}\partial_j\mid\phi\mid^2 - \phi^*\partial_i\phi + \phi\partial_i\phi^*). \qquad (77)$$

In this way, to achieve full consistency between the BPS analysis with the variational dynamical equations to $O(\theta)$, it appears that $\phi$ has to satisfy the above set of equations (76) and (77) as well. This is the disagreement we mentioned before and the situation is similar to our previously studied instance of $CP(1)$ solitons in in [11].

### X. Conclusion

We have studied the noncommutative extension of the relativistic Chern-Simons-Higgs model, in the first non-trivial order in $\theta$. Both Lagrangian and Hamiltonian formulations of the problem have been discussed, with the focus being on the canonical and symmetric forms of the energy-momentum tensor.

In the Hamiltonian scheme, constraint analysis and the induced Dirac brackets have been provided. The spacetime translation generators and their actions on the fields are discussed in detail.

The BPS soliton solutions are obtained from the energy expression, derived from the symmetric energy-momentum tensor in the Lagrangian framework. It is shown that, in terms of phase space variables, the energy expressions obtained from canonical and symmetric forms of the energy-momentum tensor are identical on the constraint surface.

We have studied the effects of noncommutativity on the soliton solutions thoroughly and have come up with some interesting observations. Considering the *relative* strength of the noncommutative effects, we have shown that there is a universal character in the noncommutative correction to the magnetic field - it depends *only* on $\theta$. On the other hand, in the cases of all other observables of physical interest, such as the potential profile, soliton mass or the electric field, $\theta$ as well as $\tau$, (comprising solely of commutative Chern-Simons-Higgs model parameters), appear on similar footings. This phenomenon is a new finding which has come up in the present analysis.

Lastly, we have pointed out a generic problem in the NC extension of the models, in the form of a mismatch between the BPS dynamical equation and the full variational equations of motion, to $O(\theta)$. This indicates that although the existence of the solitons in nocommutative Chern-Simons-Higgs model model or in nocommutative $CP(1)$ model [11] is well established, with the soliton solutions being well behaved having a smooth $\theta = 0$ limit, the analysis is not complete. This is because the above mentioned mismatch brings in to fore the ambiguities in the definition of the energy-momentum tensor in a noncommutative theory.

# References

[1] S.Deser and R.Jackiw, Phys.Lett. 139B(1984)371; S.Deser, R.Jackiw and S.Templeton, Phys.Rev.Lett. 48 (1982)975; Ann.Phys. 140 (1982)372.

[2] J.Hong, Y.Kim and P.Y.Pac, Phys.Rev.Lett. 64 2230(1990); R.Jackiw and E.J.Weinberg, *ibid* 64 2234(1990).

[3] For a review, see G.Dunne, *Self-Dual Chern-Simons Theories*, Lecture Notes in Physics, Springer-Verlag, 1995.

[4] For a review, see F.Wilczek, *Fractional Statistics and Anyon Superconductivity*, World Scientific, Singapore, 1990.

[5] For reviews see for example M.R.Douglas and N.A.Nekrasov, Rev.Mod.Phys. 73(2001)977; R.J.Szabo, *Quantum Field Theory on Noncommutative Spaces*, hep-th/0109162.

[6] N.Seiberg and E.Witten, JHEP 9909(1999)032.

[7] R.Gopakumar, S.Minwalla and A.Strominger, JHEP 0005 (2000)020.

[8] A.Hashimoto and K.Hashimoto, JHEP 9911 (1999)005; D.J.Gross and N.Nekrasov, hep-th/0010090.

[9] G.S.Lozano, E.F.Moreno and F.A.Schaposnik, hep-th/0011205; D.Bak, K.Lee and J.H.Park, hep-th/0011099.

[10] B.-H.Lee, K.Lee and H.S.Yang, Phys.Lett. B498 (2001)277; K.Furuta et al., *Low energy dynamics of Noncommutative $CP^1$ Solitons in 2+1 Dimensions*, hep-th/0203125; H.Otsu et al., *New BPS Solitons in 2+1 Dimensional Noncommutative $CP^1$ Model*, hep-th/0303090.

[11] S.Ghosh, Nucl.Phys. B670 359(2003).

[12] N.Grandi and G.A Silva, Phys.Lett. 507B(2001)345.

[13] S.Ghosh, hep-th/0310155.

[14] J.M.Grimstrup et. al., hep-th/0210288; A.Das and J.Frenkel, Phys.Rev.D 67 (2003)067701 (hep-th/0212122).

[15] N.Seiberg, hep-th/0005015.

[16] B.Jurco et. al., Eur.Phys.J. C21 (2001)383.

[17] P.A.M.Dirac, *Lectures on Quantum Mechanics*, Yeshiva University Press, New York, 1964.

Figure I-A

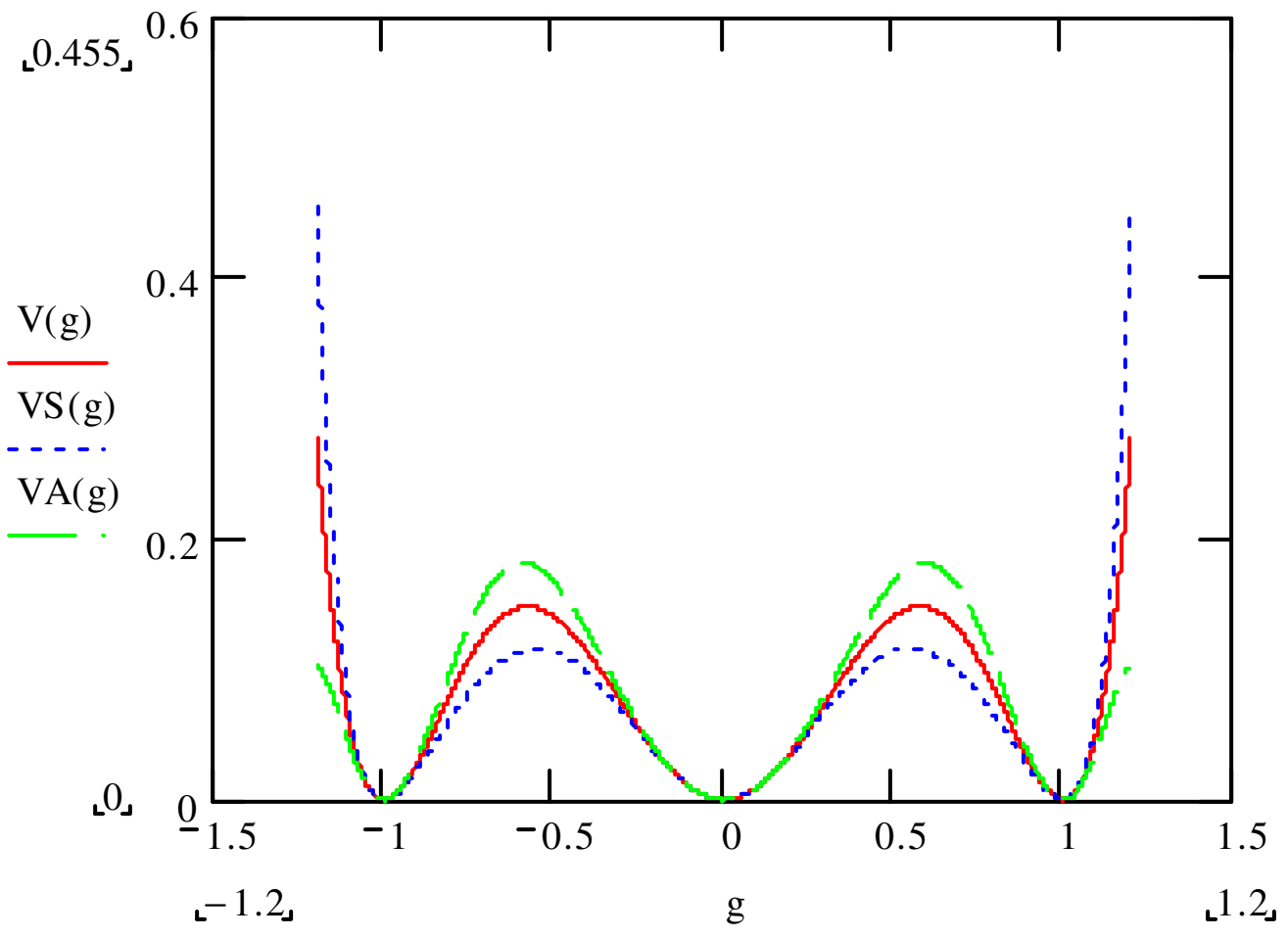

Figure I-B

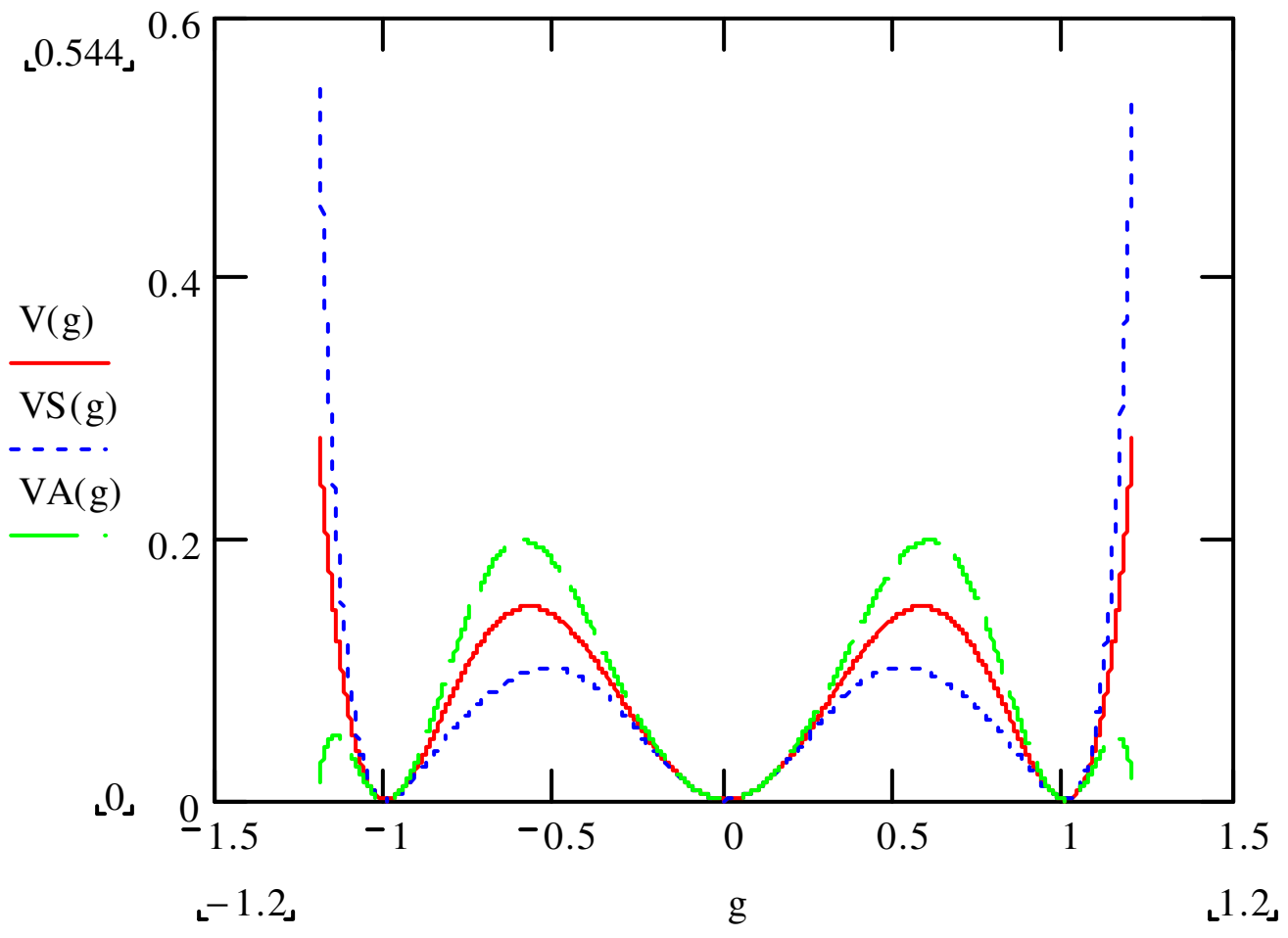

We plot the potentials VS (for self-dual) and VA (for anti-self-dual) against g for $\theta$=0.5 and $\tau$=0.5 and $\tau$=1.5 in I-A and I-B respectively. This shows that for anti-self-dual case, there is a critical value of $\tau$ above which the double well will not be formed. For the self-dual case, the potential flattens out for large $\tau$ as shown in Figure II.

Figure II

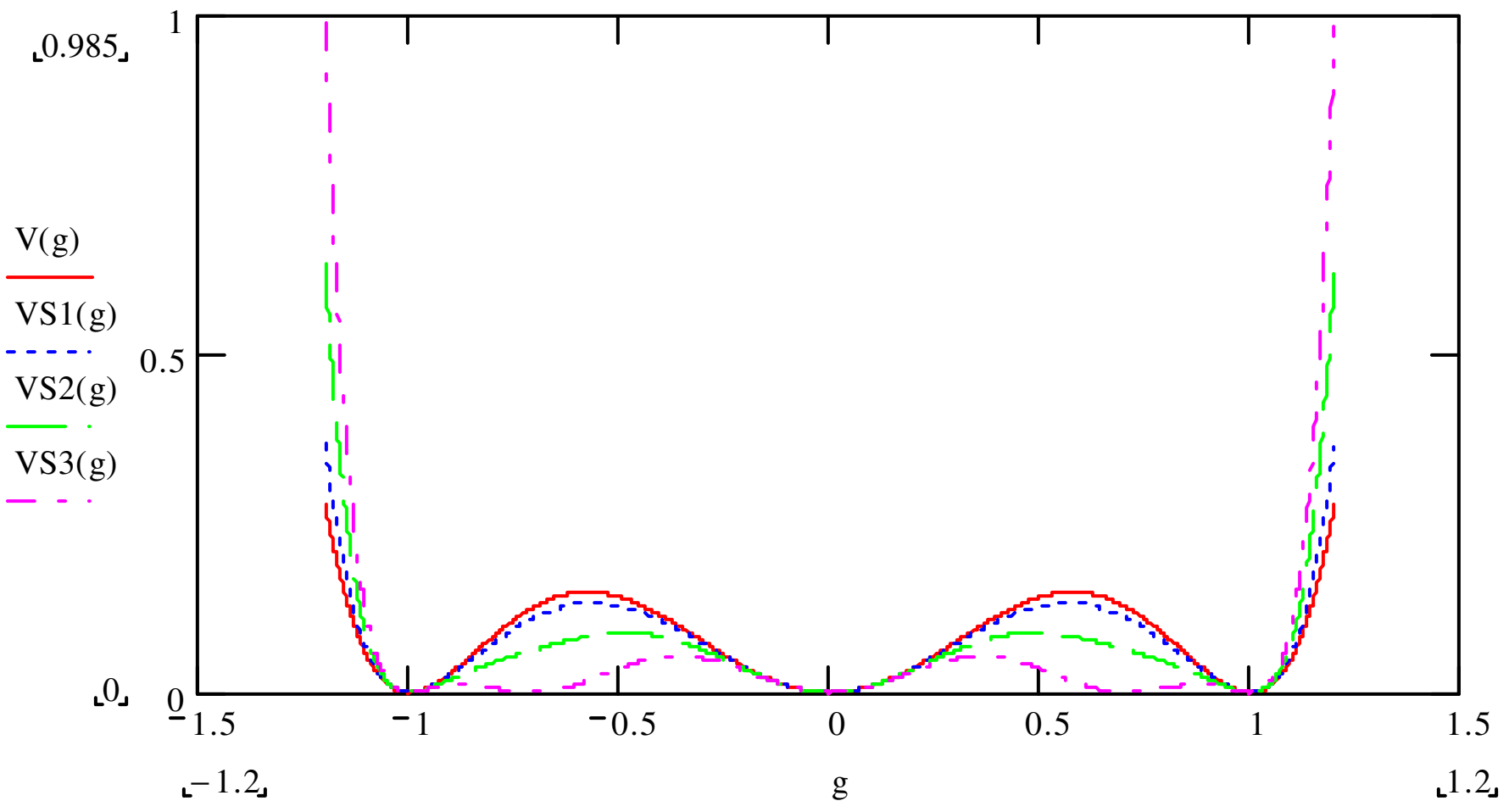

We plot the self-dual potential VS for $\theta$=0.5 and $\tau$=0.5, 2.0 and 4.0 to show the flattening of the well.

Figure III-A

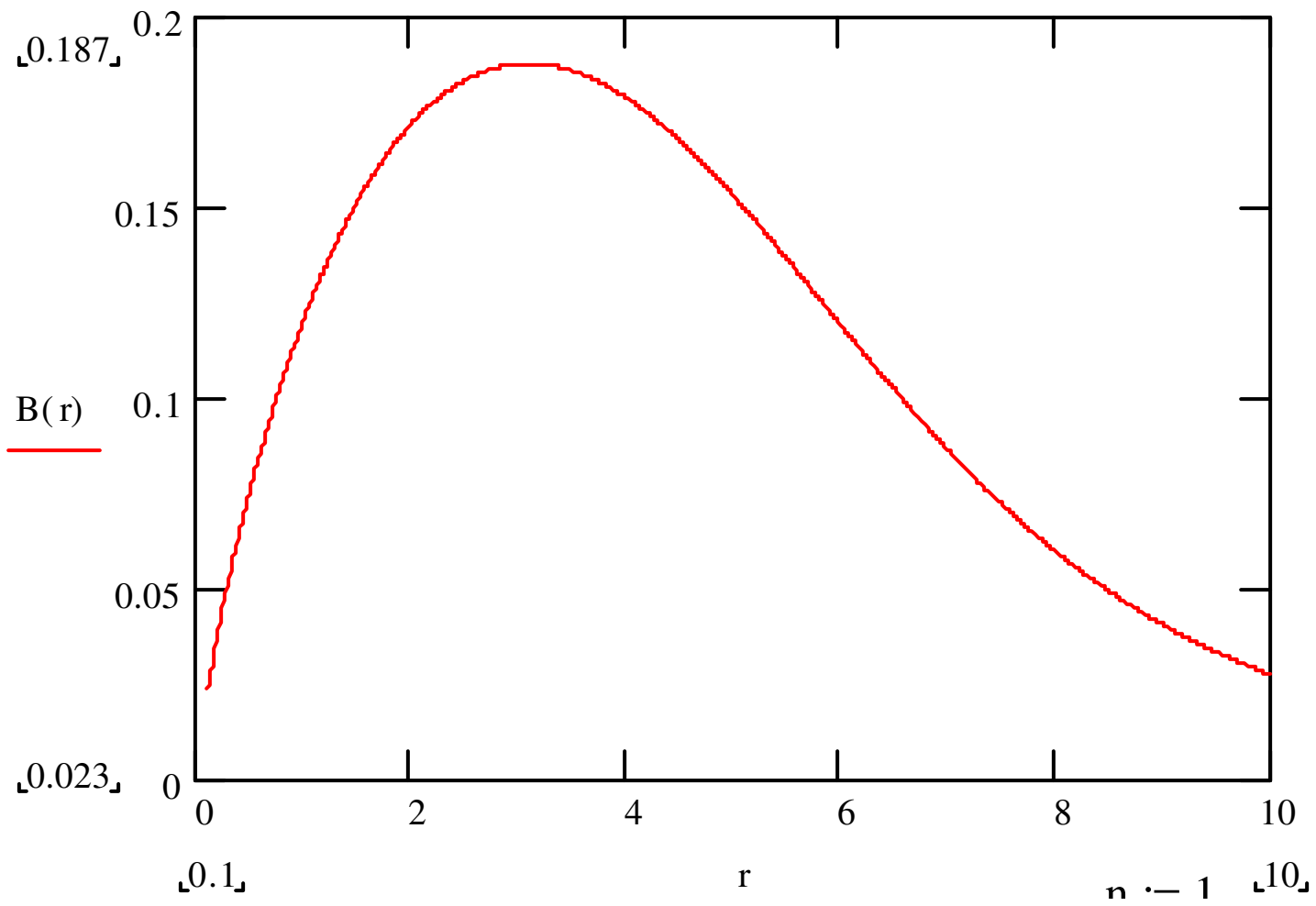

Figure III-B

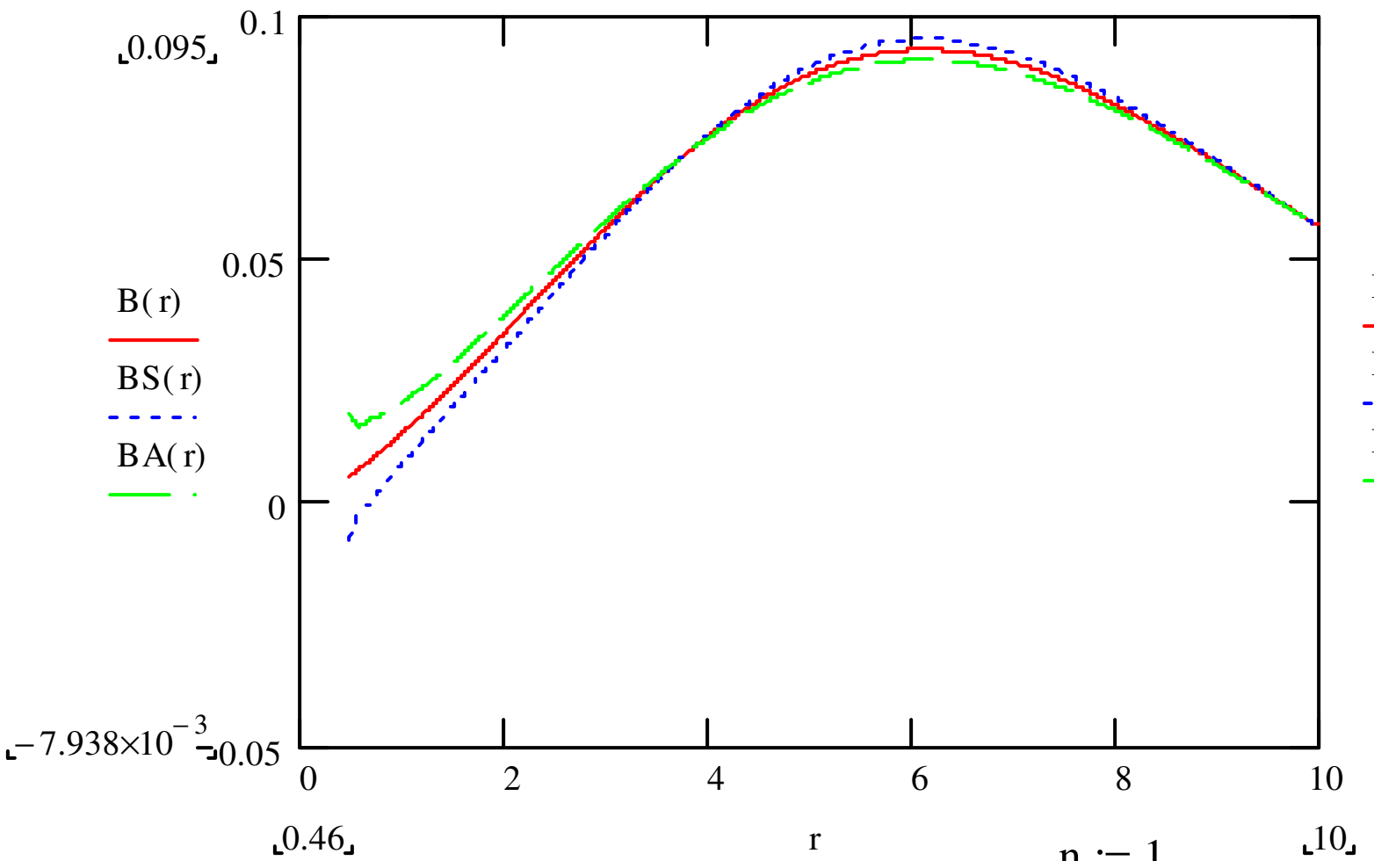

Figure III-C

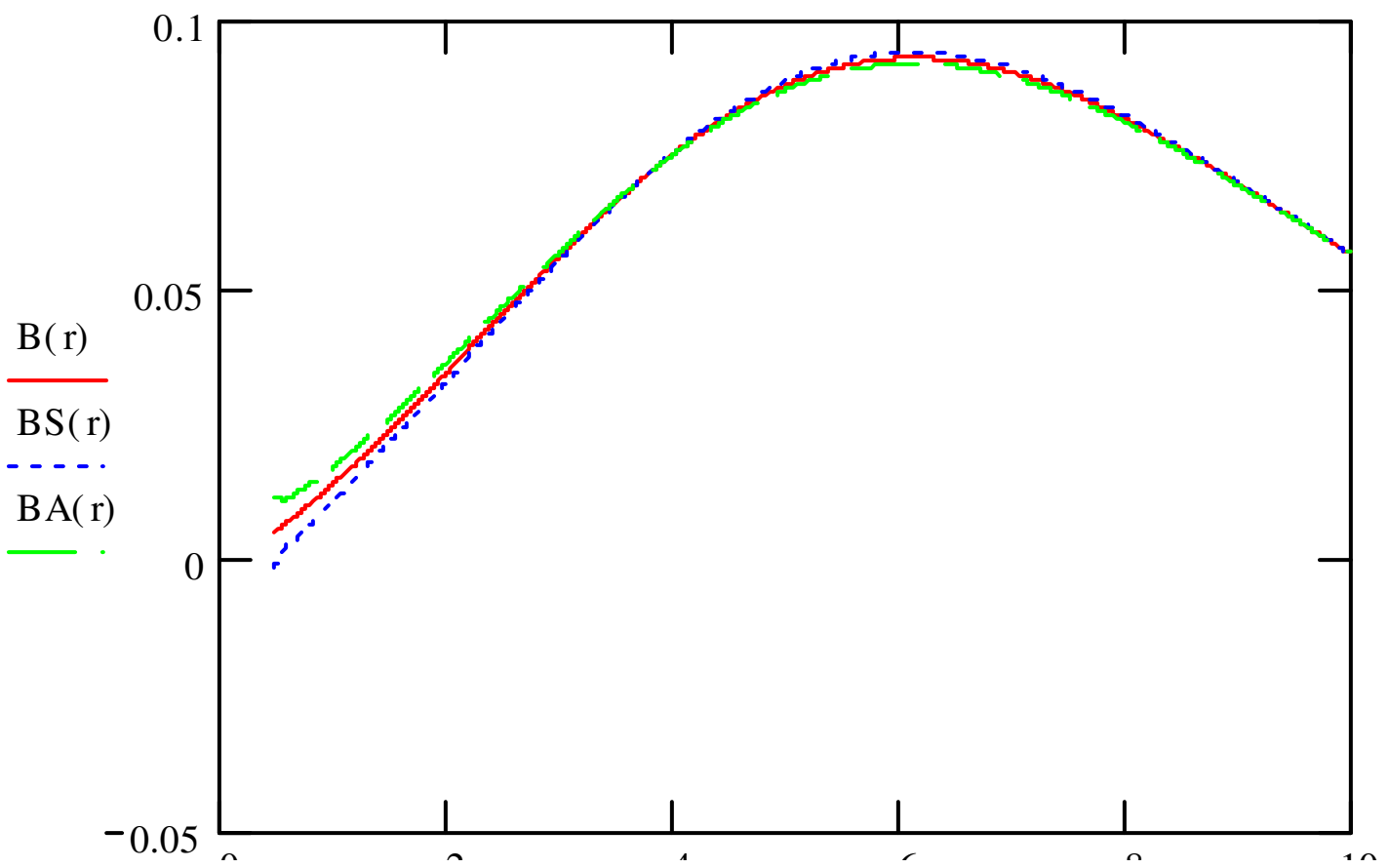

We plot the magnetic field B for θ=0 in Fig.III-A. In Figs. III-B and III-C we plot B, BS (self-dual) and BA (anti-self-dual) profiles for θ=1.0 and θ=0.5 respectively. Except for the small r region, the effect of noncommutativity is not very pronounced.

Figure IV-A

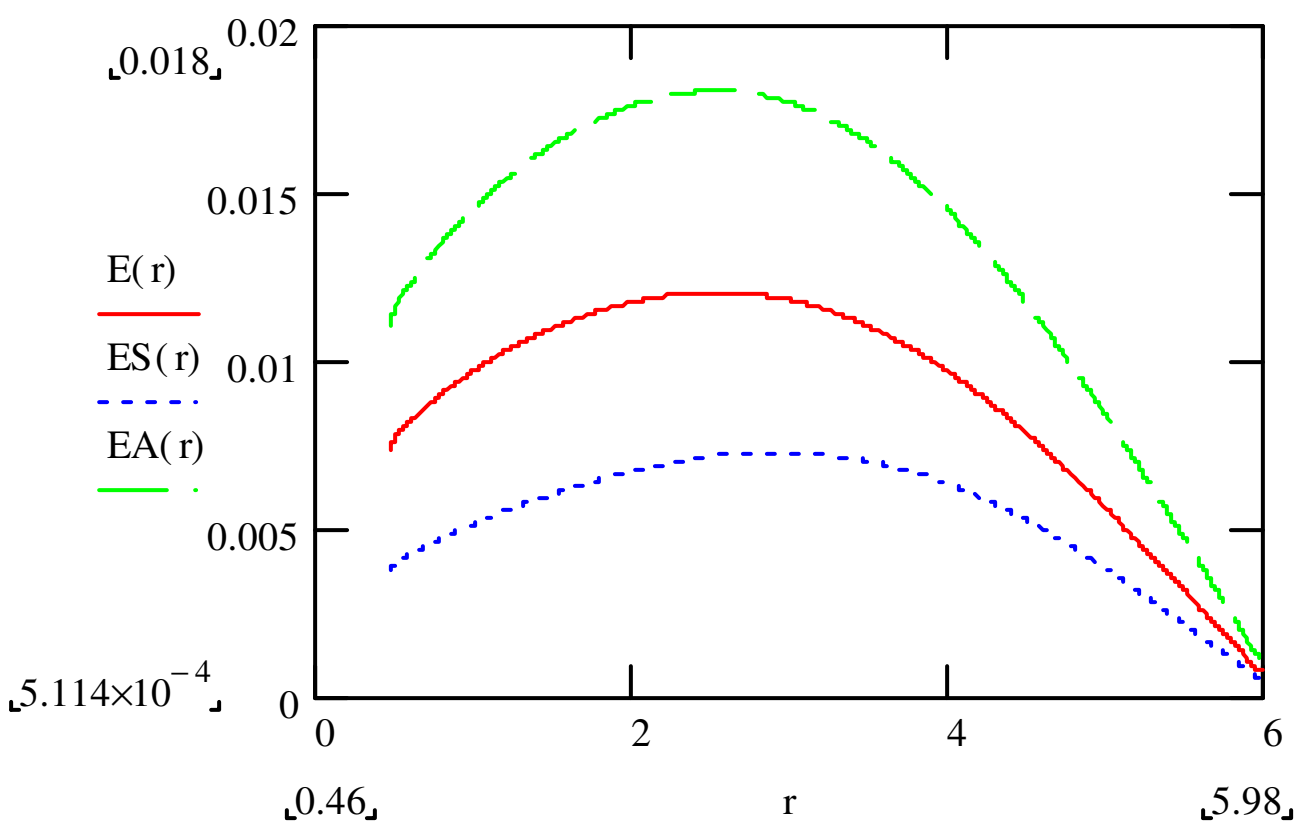

Figure IV-B

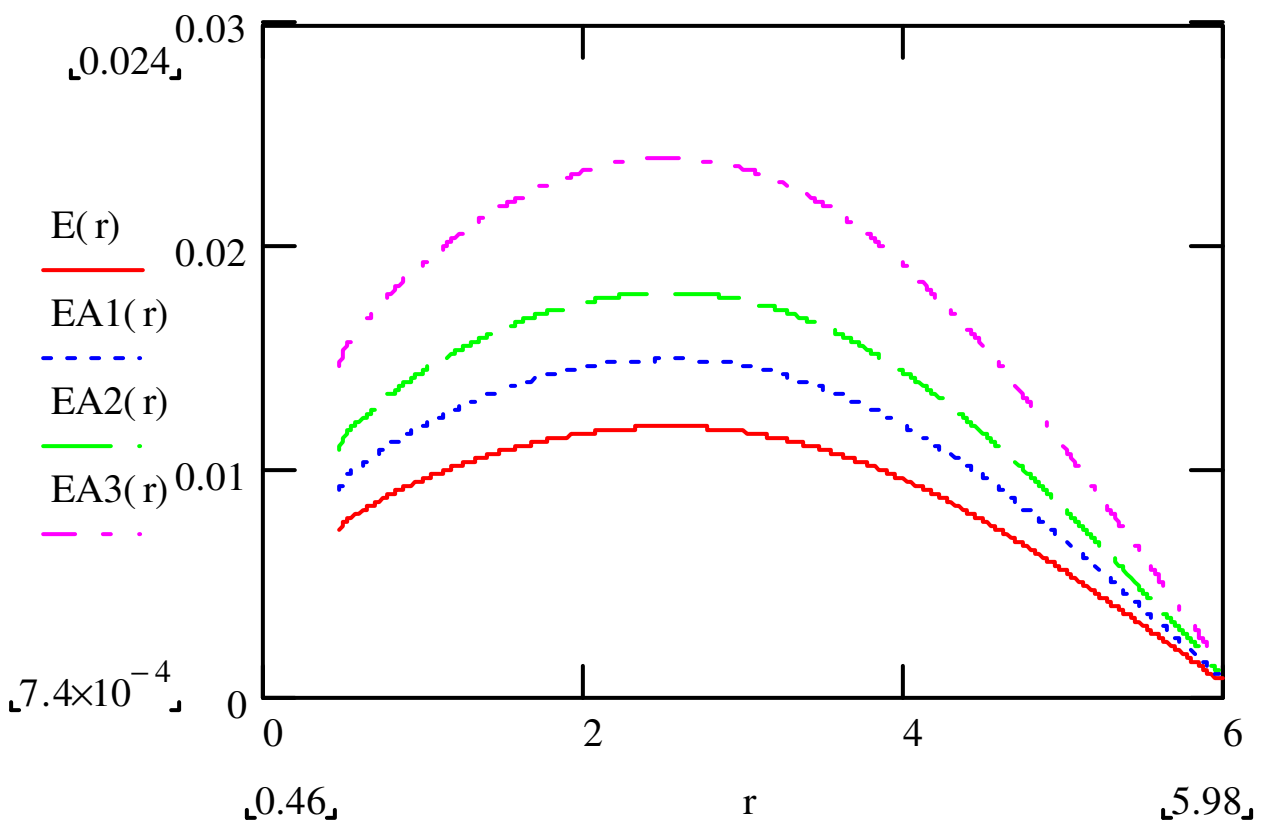

Figure IV-C

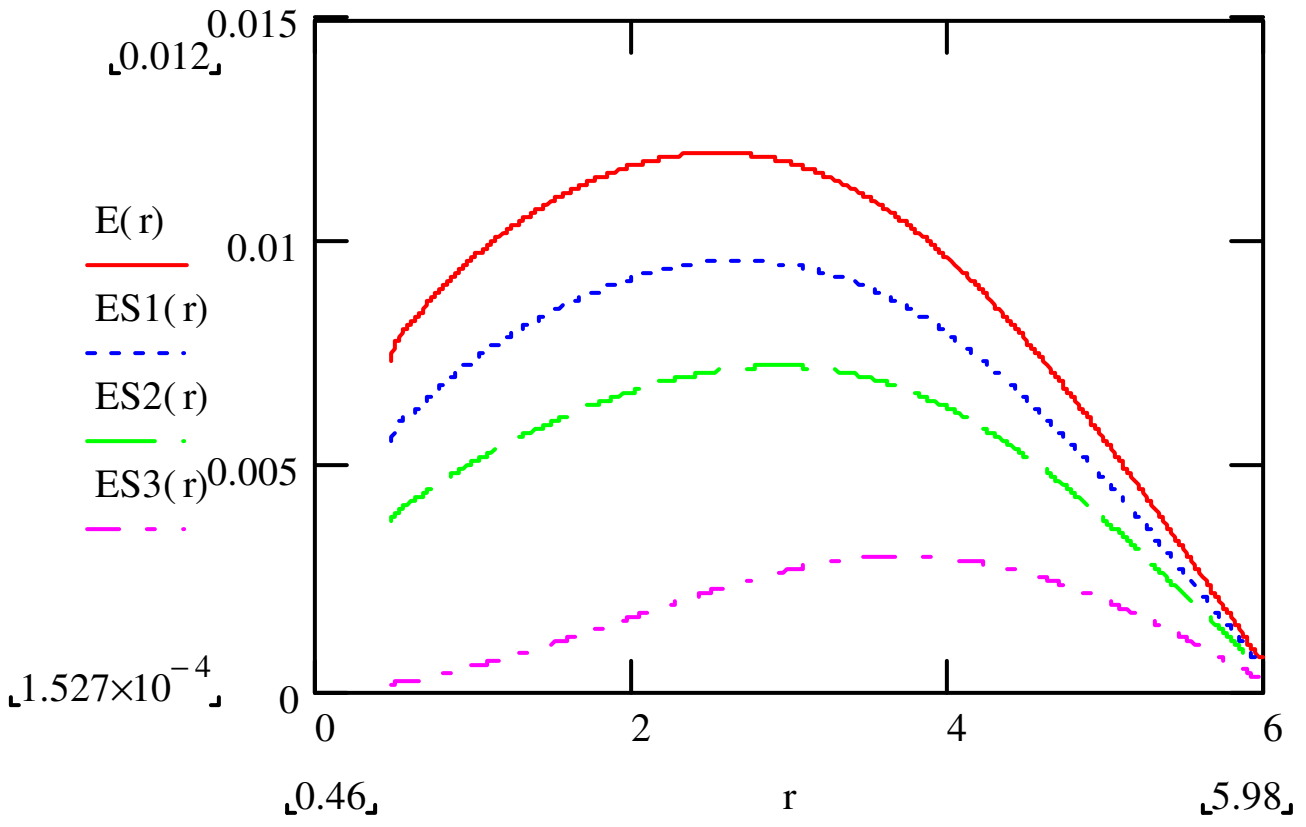

In Fig.IV-A the radial component of the Electric field ES(r) for self-dual and EA(r) anti-self-dual solutions for $\theta$=0.5 and $\tau$=0.5 are compared with the Electric field E(r) for $\theta$=0. The effect of the variation of $\tau$ are shown in the Figs. IV-B and IV-C for self-dual and anti-self-dual solutions respectively. The values of $\tau$ chosen are $\tau$=0.5, 1.0 and 2.0, for the same $\theta$=0.5.

Figure V-A

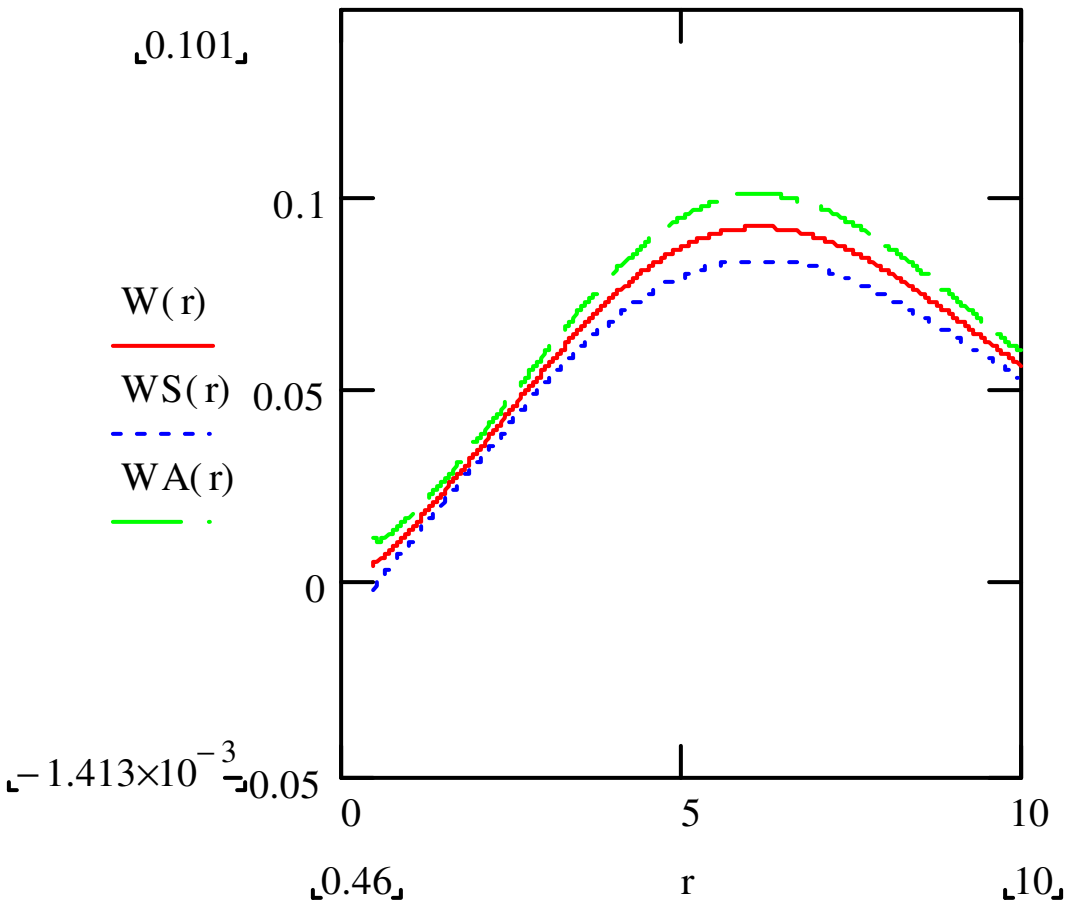

Figure V-B

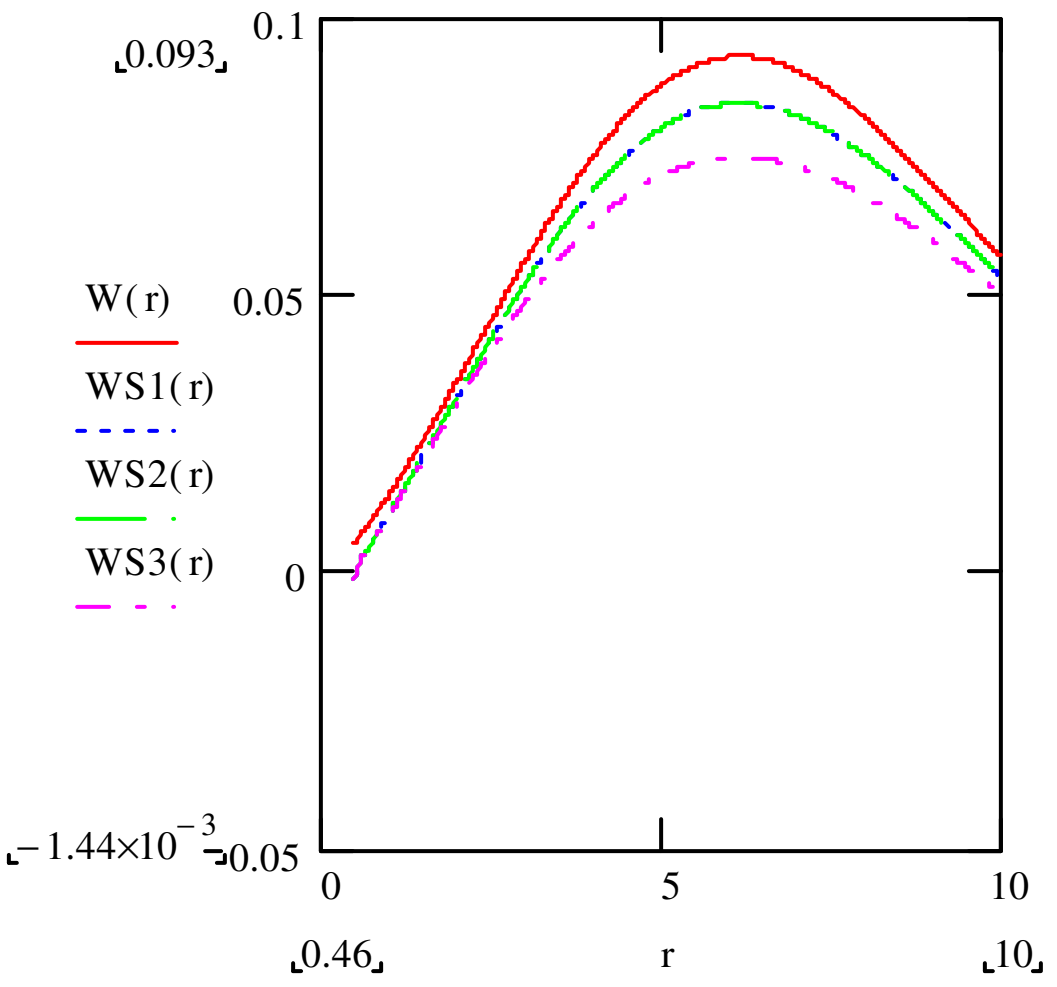

Figure V-C

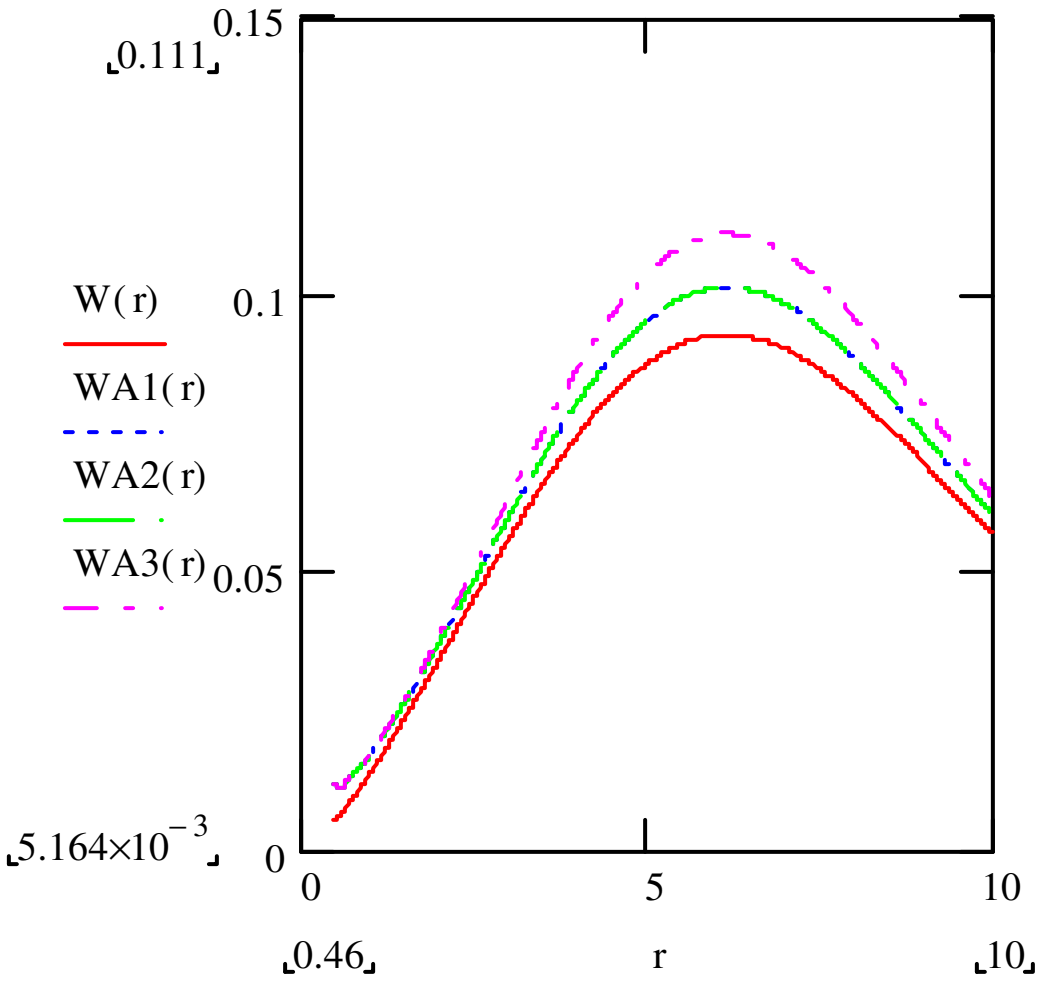

In Fig.V-A the energy density W(r) for $\theta=0$ is compared with WS(r) self-dual and WA(r) anti-self-dual solutions for $\theta=0.5$ and $\tau=0.5$. The effect of the variation of $\tau$ are shown in the Figs. V-B and V-C for self-dual and anti-self-dual solutions respectively. The values of $\tau$ chosen are $\tau=0.5$, 1.0 and 2.0, for the same $\theta=0.5$.